# INDRA: Intrusion Detection using Recurrent Autoencoders in Automotive Embedded Systems

Vipin Kumar Kukkala, *Student Member, IEEE*, Sooryaa Vignesh Thiruloga, *Student Member, IEEE*, and Sudeep Pasricha, *Senior Member, IEEE*

*Abstract*— Today's vehicles are complex distributed embedded systems that are increasingly being connected to various external systems. Unfortunately, this increased connectivity makes the vehicles vulnerable to security attacks that can be catastrophic. In this work, we present a novel Intrusion Detection System (IDS) called *INDRA* that utilizes a Gated Recurrent Unit (GRU) based recurrent autoencoder to detect anomalies in Controller Area Network (CAN) bus-based automotive embedded systems. We evaluate our proposed framework under different attack scenarios and also compare it with the best known prior works in this area.

## I. INTRODUCTION

MODERN vehicles can be considered as complex distributed embedded systems that consist of tens of interconnected Electronic Control Units (ECUs). These ECUs control various components in the vehicle and communicate with each other using the in-vehicle network. The number of ECUs and the complexity of software running on these ECUs has been steadily increasing in emerging vehicles, to support state-of-the-art Advanced Driver Assistance Systems (ADAS) features such as lane keep assist, collision warning, blind spot warning, parking assist, etc. This in turn has resulted in an increase in the complexity of the in-vehicle network, which is the backbone over which massive volumes of heterogeneous sensor and real-time decision data and control directives are communicated.

The trend in recent ADAS solutions has been to interact with external systems using advanced communication standards such as Vehicle-to-X (V2X) and 5G technology [1]. Unfortunately, this makes modern vehicles highly vulnerable to various security attacks that can be catastrophic. Several attacks have been demonstrated in [2]-[4] showing different ways to gain access to the in-vehicle network and take control of the vehicle via malicious messages. With connected and autonomous vehicles becoming increasingly ubiquitous, these security issues will only get worse. Hence, it is crucial to prevent unauthorized access of in-vehicle networks from external attackers to ensure the security of automotive systems.

Traditionally, firewalls are used to defend a network from various external attackers. *However, no firewall is perfect and no network is impenetrable*. Hence, there is a need for an active monitoring system that scans the network to detect the presence of an attacker in the system. This can be achieved using an intrusion detection system (IDS) which monitors' network traffic and triggers alerts when suspicious activity or known threats are detected. The IDS is often the last line of defense in automotive systems.

IDSs can be classified into two types: *(i) signature-based*, and *(ii) anomaly-based.* The former observes for traces of any known attack signatures while the latter observes for the deviation from the known normal system behavior to indicate the presence of an attacker. Signature-based IDS can have faster detection times and very few false positives, but can only detect known attacks. On the other hand, anomaly-based IDS can detect both known and unknown attacks, but can suffer from higher false positives and relatively slower detection times. An efficient IDS needs to be *lightweight, robust* and *scalable* with different system sizes. Moreover, a pragmatic IDS needs to have a *large coverage of attacks* (able to detect both known and unknown attacks), *high confidence in detection,* and *low false positive rate* as recovery from false positives can be expensive.

Since getting the signature of every possible attack is impractical and would limit us to only detecting known attacks, we conjecture that using anomaly-based IDS is a more practical approach to this problem. Additionally, due to the ease of in-vehicle network data acquisition (from test driving), there can be a large amount of in-vehicle message data to work with, which facilitates the use of advanced deep learning models for detecting the presence of an attacker in the system.

In this article, we propose a novel IDS framework called *INDRA* that monitors the messages in Controller Area Network (CAN) based automotive systems for the presence of an attacker. In the offline phase, *INDRA* uses deep learning to learn the normal system behavior in an unsupervised fashion. At runtime, *INDRA* monitors the network and indicates the presence of an attacker if any anomalies (any deviation from the normal behavior learned during the offline phase) are detected. *INDRA* aims to maximize the detection accuracy and minimize false positive rate with minimal overhead on the ECUs.

Our novel contributions in this work are as follows:
- We propose a Gated Recurrent Unit (GRU) based recurrent

autoencoder network to learn the latent representation of normal system behavior during the offline phase;
- We propose a metric called intrusion score (IS), which is a measure of deviation from the normal system behavior;
- We perform a thorough analysis towards the selection of thresholds for this intrusion score metric;
- We compare our proposed *INDRA* framework with the best known prior works in the area, to show its effectiveness.

## II. RELATED WORK

Several techniques have been proposed to design IDS for time-critical automotive systems. The goal of these works is to detect various types of attacks by monitoring the in-vehicle network traffic.

Signature-based IDS relies on detecting known and pre-modeled attack signatures. The authors in [5], used a language theory-based model to derive attack signatures. However, this technique fails to detect intrusions when it misses the packets transmitted during the early stages of an attack. In [6], the authors used transition matrices to detect intrusions in a CAN bus. In spite of achieving a low false-positive rate for trivial attacks, this technique failed to detect realistic replay attacks. The authors in [7] identify notable attack patterns such as an increase in message frequency and missing messages to detect intrusions. A specification-based approach to detect intrusions is proposed in [8], where the authors analyze the behavior of the system and compare it with the predefined attack patterns to detect intrusions. Nonetheless, their system fails to detect unknown attacks. In [9], an IDS technique using the Myers algorithm [10] was proposed under the map-reduce framework. A time-frequency analysis of CAN messages is used in [11] to detect multiple intrusions. A rule-based regular operating mode region is derived in [12] by analyzing the message frequency at design time. This region is observed for deviations at runtime to detect anomalies. In [13], fingerprints of the sender ECU's clock skew and the messages are used to detect intrusions by observing for variations in the clock-skew at runtime. In [14], a formal analysis is presented for clock-skew based IDS and evaluated on a real vehicle. A memory heat map is used to characterize the memory behavior of the operating system to detect intrusions in [15]. An entropy-based IDS is proposed in [16] that observes for change in system entropy to detect intrusions. However, the technique fails to detect small scale attacks where the entropy change is minimal. *In summary, signature-based techniques offer a solution to the intrusion detection problem with low false positive rates but cannot detect more complex and novel attacks. Moreover, modeling signatures of every possible attack is impractical.*

Anomaly-based IDS aims to learn the normal system behavior in an offline phase and observe for any deviation from the learned normal behavior to detect intrusions at runtime. A sensor-based IDS was proposed in [17], that utilizes attack detection sensors to monitor various system events to observe for deviations from normal behavior. However, this approach is not only expensive but also suffers from poor detection rates. A One-Class Support Vector Machine (OCSVM) based IDS was proposed in [18]. However, this approach suffers from poor detection latency. In [19], the authors used four different nearest neighbor classifiers to distinguish between a normal and an attack induced CAN payload. In [20], a decision-tree based detection model is proposed to monitor the physical features of the vehicle to detect intrusions. However, this model is not realistic and suffers from high detection latencies. A Hidden Markov Model (HMM) based technique was proposed in [21] that monitors the temporal relationships between messages to detect intrusions. A deep neural network based approach was proposed to examine the messages in the in-vehicle network in [22]. This approach is tuned for a low priority tire pressure monitoring system (TPMS), which makes it hard to adapt to high priority safety-critical powertrain applications. A Long Short-Term Memory (LSTM) based IDS for multi-message ID detection was proposed in [23]. However, the model architecture is highly complex, which incurs high overhead on the ECUs. In [24], the authors use LSTM based IDS to detect insertion and dropping attacks (explained in section IV C). An LSTM based predictor model is proposed in [25] that predicts the next time step message value at a bit level and observes for large variations in loss to detect intrusions. In [26], a recurrent neural network (RNN) based IDS was proposed to learn the normal patterns in CAN messages in the in-vehicle network. A hybrid IDS was proposed in [27], which utilizes a specification-based system in the first stage and an RNN based model in the second stage to detect anomalies in time-series data. *However, none of these techniques provides a holistic system-level solution that is lightweight, scalable, and reliable to detect multiple types of attacks for in-vehicle networks.*

In this paper, we propose a lightweight recurrent autoencoder based IDS using gated recurrent units (GRUs) that monitors messages at a signal level granularity to detect multiple types of attacks more effectively and successfully than the state of the art. Table I summarizes some of the state-of-the-art works' performance under different metrics and shows how our proposed *INDRA* framework fills the existing research gap. The *INDRA* framework aims at improving multiple performance metrics compared to the state-of-the art IDS works that target a subset of performance metrics. A detailed analysis of each metric and evaluation results are presented later in section VI.

TABLE I
COMPARISON BETWEEN PROPOSED *INDRA* FRAMEWORK
AND STATE-OF-THE-ART WORK

| Technique | Performance metrics | | | |
|---|---|---|---|---|
| | Lightweight | Low False Positive Rate | High accuracy | Fast Inference |
| PLSTM [25] | X | ✓ | X | X |
| RepNet [26] | ✓ | X | X | ✓ |
| CANet [23] | X | ✓ | ✓ | X |
| INDRA | ✓ | ✓ | ✓ | ✓ |

## III. SEQUENCE LEARNING BACKGROUND

With the availability of increased computing power from GPUs and custom accelerators, training neural networks with many hidden layers (known as *deep neural networks*) has led to the creation of powerful models for solving difficult problems in many domains. One such problem is detecting intrusions in the in-vehicle network. In an in-vehicle network, the communication between ECUs happens in a time-dependent manner. Hence, there exist temporal relationships between the messages, which is essential to exploit, in order to detect intrusions. However, this cannot be achieved using traditional feedforward neural networks where the output of any input is independent of the other inputs. Sequence models are a more appropriate approach for such problems, as they are designed to handle sequences and time-series data.

*A. Sequence Models*

A sequence model can be understood as a function which ensures that the output is dependent not only on the current input, but also on the previous inputs. An example of such a sequence model is the recurrent neural network (RNN), which was introduced in [28]. In recent years, other sequence models such as long short-term memory (LSTM) and gated recurrent unit (GRU) have also been developed.

*1) Recurrent Neural Networks (RNN)*

An RNN is a type of neural network which takes sequential data as the input and learns the relationship between data sequences. RNNs have hidden states, which allows learned information to persist over time steps. The hidden states enable the RNN to connect previous information to current inputs. An RNN cell with feedback is shown in Fig. 1a, and an unrolled RNN in time is shown in Fig. 1b.

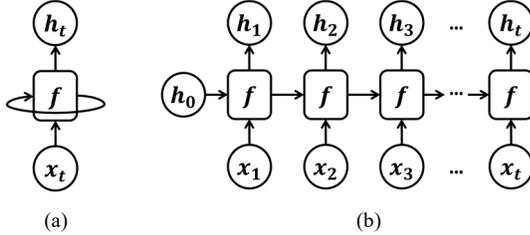
Figure 1: (a) A single RNN cell and (b) unrolled RNN unit; where, f is the RNN cell, x is the input, and h represents hidden states.

The output $h_t$ of an RNN cell is a function of both the input $x_t$ and the previous output $h_{t-1}$:

$$h_t = f(Wx_t + Uh_{t-1} + b) \quad (1)$$

where $W, U$ are weight matrices, $b$ is a bias term, and $f$ is a nonlinear activation function (e.g. sigmoid or tanh). One of the limitations of RNNs is that they are very hard to train. Since RNNs and other sequence models deal with sequence or time-series inputs, backpropagation happens through various time samples (known as backpropagation through time). During this process, the feedback loop in RNNs causes the errors to shrink or grow rapidly (creating vanishing or exploding gradients respectively), destroying the information in backpropagation. This problem of vanishing gradients hampers the RNNs from learning *long term dependencies*. This problem was solved in [29] with the introduction of additional states and gates in the RNN cell to remember long term dependencies, which led to the introduction of Long Short-Term Memory Networks.

*2) Long Short Term Memory (LSTM) Networks*

LSTMs are modified RNNs that use cell state and hidden state information along with multiple gates to remember long term dependencies. The cell state can be thought of as a transport highway, that carries relevant information throughout the processing of a sequence. The state accommodates the information from earlier time steps, which can be used in the later time steps, thereby reducing the effects of short-term memory. The information in the cell state is modified via gates. Hence, the gates in LSTM help the model decide which information has to be retained and which information to forget.

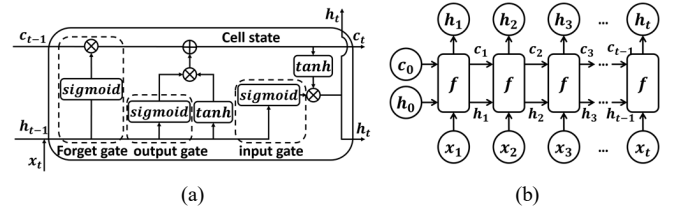
Figure 2: (a) A single LSTM cell with different gates and (b) unrolled LSTM unit; where, $f$ is an LSTM cell, $x$ is input, $c$ is cell state and $h$ is the hidden state.

An LSTM cell consists of 3 gates: *(i)* forget gate ($f_t$) *(ii)* input gate ($i_t$), and *(iii)* output gate ($o_t$) as shown in Fig.2(a). The forget gate is a binary gate that chooses which information to retain from the previous cell state ($c_{t-1}$). The input gate adds relevant information to the cell state ($c_t$). Lastly, the output layer is controlled by the output gate, which uses information from the previous two gates to produce an output. An unrolled LSTM unit is shown in Fig. 2b.

By using the combination of different gates and hidden states, LSTMs can learn long term dependencies in a sequence. However, they are not computationally efficient as the sequence path is more complicated than in RNNs, due to the addition of multiple gates, requiring more memory at runtime. Moreover, training LSTMs is compute intensive even with advanced training methods such as truncated backpropagation. To overcome these limitations, a simpler recurrent neural network called gated recurrent unit (GRU) network was introduced in [30] that can be trained faster than LSTMs and also remembers dependencies in long sequences with low memory overhead, while solving the vanishing gradient problem.

*3) Gated Recurrent Unit (GRU)*

A GRU cell uses an alternate route for gating information when compared to LSTMs. It combines the input and forget gate of the LSTM into a solitary *update gate* and furthermore combines hidden and cell state, as shown in Fig 3a and 3b.

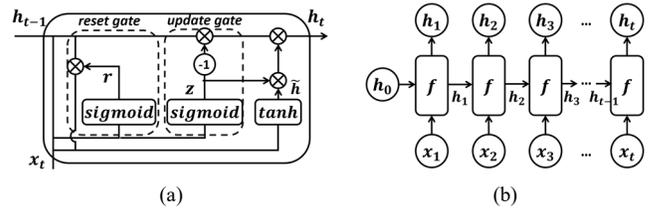
Figure 3: (a) A single GRU cell with different gates and (b) unrolled GRU unit; where, f is a GRU cell, x is input, and h represents hidden states.

A traditional GRU cell has two gates (i) reset gate and (ii) update gate. The reset gate combines new input with past memory while the update layer chooses the amount of pertinent data that should be held. Thus, a GRU cell can control the data stream like an LSTM by uncovering its hidden layer contents. Moreover, GRUs achieve this using fewer gates and states, which makes them computationally more efficient with low memory overhead. As real-time automotive ECUs are highly resource-constrained embedded systems with tight energy and power budgets, it is critical to use low overhead models for inferencing tasks. Thus, GRU based networks are an ideal fit for inference in automotive systems. Additionally, GRUs are relatively new, less explored and have a lot of potential to offer compared to the RNNs and LSTMs. Hence, in this work, we chose to use a lightweight GRU based model to implement our IDS (explained in detail in section V).

One of the advantages of sequence models is that they can be trained using both supervised and unsupervised learning approaches. As there is a large volume of CAN message data in a vehicle, labeling the data can become very tedious. Additionally, the variability in the messages between vehicle models from the same manufacturer and the proprietary nature of this information, makes it even more challenging to label messages correctly. However, due to the ease of availability to CAN message data via onboard diagnostics (OBD-II), large amounts of unlabeled data can be collected easily. Thus, we use GRUs in an unsupervised learning setting in this work.

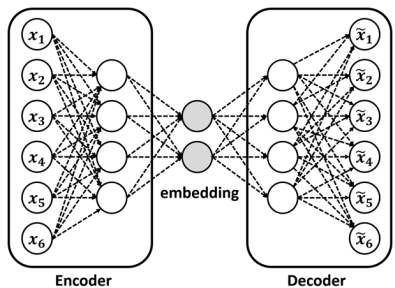

Figure 4: Autoencoders

### B. Autoencoders

An autoencoder is an unsupervised learning algorithm whose goal is to reconstruct the input by learning latent input features. It achieves this by encoding the input data ($x$) towards a hidden layer, and finally decodes it to produce a reconstruction $\tilde{x}$ (as shown in Fig. 4). This encoding at the hidden layer is called an embedding. The layers that create this embedding are called the encoder, and the layers that reconstruct the embedding into the original input are called the decoder. During training, the encoder tries to learn a nonlinear mapping of the inputs, while the decoder tries to learn the nonlinear mapping of the embedding to the inputs. Both encoder and decoder achieve this with the help of non-linear activation functions, e.g., tanh, and relu. Moreover, the autoencoder aims to recreate the input as accurately as possible by extracting the key features from the inputs with a goal of minimizing reconstruction loss. The most commonly used loss functions in autoencoders are mean squared error (MSE) and Kullback-Leibler (KL) divergence.

As autoencoders aim to reconstruct the input by learning the underlying distribution of the input data, it makes them an ideal choice to learn and reconstruct highly correlated time-series data efficiently by learning the temporal relations between signals. *Hence, our proposed INDRA framework uses light weight GRUs in an autoencoder to learn latent representations of CAN messages in an unsupervised learning setting.*

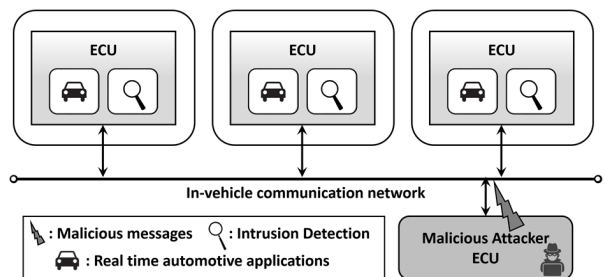

Figure 5: Overview of the system model

## IV. PROBLEM DEFINITION

### A. System Model

We consider a generic automotive system consisting of multiple ECUs connected using a CAN based in-vehicle network, as shown in Fig. 5. Each ECU is responsible for running a set of automotive applications that are hard-real time in nature, meaning they have strict timing and deadline constraints. In addition, we assume that each ECU also executes intrusion detection applications that are responsible for monitoring and detecting intrusions in the in-vehicle network. We consider a distributed IDS approach (intrusion applications collocated with automotive applications) as opposed to a centralized IDS approach where one central ECU handles all intrusion detection tasks due to the following reasons:

- A centralized IDS approach is prone to single-point failures, which can completely open up the system to the attacker.
- In extreme scenarios such as during a flooding attack (explained in section IV-C), the in-vehicle network can get highly congested and the centralized system might not be able to communicate with the victim ECUs.
- If an attacker succeeds in fooling the centralized IDS ECU, attacks can go undetected by the other ECUs, resulting in compromising the entire system; whereas with a distributed IDS, fooling multiple ECUs is required which is much harder, and even if an ECU is compromised, this can still be detected by the decentralized intelligence.
- In a distributed IDS, ECUs can stop accepting messages as soon as an intrusion is detected without waiting for a centralized system to notify them, leading to faster response.
- The computation load of IDS is split among the ECUs with a distributed IDS, and the monitoring can be limited to only the required messages. Thus, multiple ECUs can monitor a subset of messages independently, with lower overhead.

Many prior works, e.g., in [5] and [12], consider a distributed IDS approach for these reasons. Moreover, with automotive ECUs becoming increasingly powerful, the collocation of IDS applications with real-time automotive applications in a distributed manner should not be a problem, provided the overhead from the IDS is minimal. Our proposed framework is not only lightweight, but also scalable, and achieves high intrusion detection performance, as discussed in Section VI.

The design of an IDS should have low susceptibility to noise, low cost, and a low power/energy footprint. The following are some of the goals that we considered for our IDS:

- *Lightweight*: Intrusion detection tasks can incur overhead on

the ECUs that could result in poor application performance or missed deadlines for real-time applications. This can be catastrophic in some cases. Hence, we aim to have a lightweight IDS that incurs low overhead on the system.
- *Few false positives*: This is a highly desired quality in any kind of IDS (even outside of the automotive domain), as handling false positives can become expensive very quickly. A good IDS needs to have few false positives or false alarms.
- *Coverage*: This is the range of attacks an IDS can detect. A good IDS needs to be able to detect more than one type of attack. A high coverage for IDS will make the system resilient to multiple attack surfaces.
- *Scalability*: This is an important requirement as emerging vehicles have increasing numbers of ECUs, and high software and network complexity. A good IDS should be highly scalable and be able to support multiple system sizes.

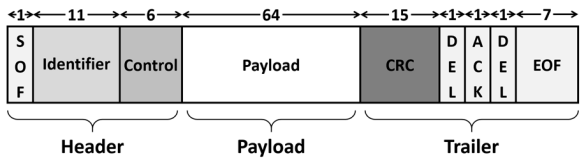

Figure 6: Standard CAN frame format

### B. Communication Model

In this subsection, we discuss the vehicle communication model that was considered. We primarily focus on detecting intrusions in a CAN bus-based automotive system. Controller Area Network (CAN) is the defacto industry standard in-vehicle network protocol for automotive systems today. CAN is a lightweight, low cost, and event-triggered communication protocol that transmits messages in the form of frames. The structure of a standard CAN frame is shown in Fig. 6 and the length of each field (in bits) is shown on the top. The standard CAN frame consists of header, payload, and trailer segments. The header consists of information such as the message identifier (ID) and the length of the message. The actual data that needs to be transmitted is in the payload segment. The trailer section is mainly used for error checking at the receiver. A variation of the standard CAN, called CAN-extended or CAN 2.0B is also becoming increasingly common in modern vehicles. The major difference is that CAN extended has a 29-bit identifier allowing for more number of messages IDs.

| Signal Name | Message | Start bit | Length | Byte Order | Value Type |
|---|---|---|---|---|---|
| Battery_Current | Status | 0 | 16 | Intel | Signed |
| Battery_Voltage | Status | 16 | 16 | Intel | Unsigned |
| Motor_Current | Status | 32 | 16 | Intel | Signed |
| Motor_Speed | Status | 48 | 8 | Intel | Signed |
| Motor_Direction | Status | 56 | 8 | Intel | Unsigned |

Figure 7: Real-world CAN message with signal information

In this work, we design our IDS with a focus on monitoring the message payload and observe for anomalies to detect intrusions. This is because an attacker needs to modify the message payload to accomplish a malicious activity. While an attacker could target the header or trailer segments, it would result in the message getting rejected at the receiver. The payload segment consists of multiple data entities called signals. An example real-world CAN message with the signals is shown in Fig. 7 [31]. Each signal has a fixed size (in bits), a particular data type, and a start bit that specifies its location in the 64-bit payload segment of the CAN message.

In this work, we focus on monitoring individual signals within message payloads to observe for anomalies and detect intrusions. Our model learns the temporal dependencies between the messages at a signal level during training and observes for deviations at runtime to detect intrusions. Signal level monitoring would give us the capability to not only detect the presence of an intruder but also helps in identifying the signal within the message that is being targeted during an attack. This can be valuable information for understanding the intentions of the attacker, which can be used for developing countermeasures. The details about the signal level monitoring of our IDS are discussed in section V-B. <u>Note</u>: Even though in this work we focus on detecting intrusions by monitoring CAN messages, our approach is protocol-agnostic and can be used with other in-vehicle network protocols.

### C. Attack Model

Our proposed IDS aims to protect the vehicle from multiple types of attacks listed below. These are some of the most common and hard to detect attacks, and have been widely considered in literature to evaluate IDS models.

*(1) Flooding attack*: This is the most common and easy to launch attack and requires no knowledge about the system. In this attack, the attacker floods the in-vehicle network with a random or specific message and prevents the other ECUs from communicating. These attacks are generally detected and prevented by the bridges and gateways in the in-vehicle network and often do not reach the last line of defense (the IDS). However, it is important to consider these attacks as they can have a severe impact when not handled correctly.

*(2) Plateau attack*: In this attack, an attacker overwrites a signal value with a constant value over a period of time. The severity of this attack depends on the magnitude of the jump (increase in signal value) and the duration for which it is held. Large jumps are easier to detect compared to shorter jumps.

*(3) Continuous attack*: In this attack, an attacker slowly overwrites the signal value with the goal of achieving some target value and avoid triggering of an IDS in the system. This attack is hard to detect and can be sensitive to the IDS parameters (discussed in section V-B).

*(4) Suppress attack*: In this attack, the attacker suppresses the signal value(s) by either disabling the communication controller of the target ECU or by powering off the ECU. These attacks can be easily detected, as they shut down message transmission for long durations, but are harder to detect for shorter durations.

*(5) Playback attack*: In this attack, the attacker replays a valid series of message transmissions from the past trying to trick the IDS. This attack is hard to detect if the IDS does not have the ability to capture the temporal relationships between messages.

In this work, we assume that the attacker can gain access to the vehicle using the most common attack vectors, which include connecting to V2X systems that communicate with the outside world (such as infotainment and connected ADAS systems), connecting to the OBD-II port, probe-based snooping on the in-vehicle bus, and via replacing an existing ECU. We also assume that the attacker has access to the bus parameters (such as BAUD rate, parity, flow control, etc.) that can help in gaining access to the in-vehicle network.

*Problem objective*: The goal of our work is to implement a lightweight IDS that can detect multiple types of attacks (as mentioned above) in a CAN based automotive system, with a high detection accuracy and low false positive rate, and while maintaining a large attack coverage.

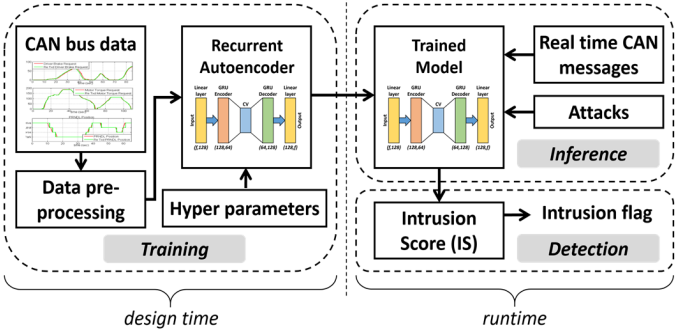

Figure 8: Overview of proposed *INDRA* framework

## V. INDRA FRAMEWORK OVERVIEW

We propose the *INDRA* framework to enable a signal level anomaly-based IDS for monitoring CAN messages in automotive embedded systems. An overview of the proposed framework is shown in Fig. 8. At a high level, the *INDRA* framework consists of design-time and runtime components. At design time, *INDRA* uses trusted CAN message data to train a recurrent autoencoder based model to learn the normal behavior of the system. At runtime, the trained recurrent autoencoder model is used for observing deviations from normal behavior (inference) and detect intrusions based on the deviation computed using the proposed intrusion score metric (detection). The following subsections describe these steps in more detail.

### A. Recurrent Autoencoder

Recurrent autoencoders are powerful neural networks that are designed to behave like an encoder-decoder but handle time-series or sequence data as inputs. They can be visualized as regular feed-forward neural network based autoencoders, with the neurons being RNN, LSTM or GRU cells (discussed in section III). Similar to regular autoencoders, the recurrent autoencoders have an encoder and a decoder stage. The encoder is responsible for generating a latent representation of the input data in an n-dimensional space. The decoder uses the latent representation from the encoder and tries to reconstruct the input data with minimal error. In this work, we propose a new lightweight recurrent autoencoder model, that is customized for the design of IDS to detect intrusions in the in-vehicle network data. The details of the proposed model architecture and the stages involved in its training are discussed next.

#### 1) Model Architecture

The proposed recurrent autoencoder model architecture with the dimensions (input, output) of each layer is illustrated in Fig. 9. The model consists of a linear layer at the input, GRU based encoder, GRU based decoder and a final linear layer before the output. The input to the first linear layer is the time-series of CAN message data with signal level values with $f$ features (where $f$ is the number of signals in that particular message). The output of the linear layer is given to the GRU based encoder to generate the latent representation of the time-series signal inputs. We call this latent representation as a message context vector (MCV). The MCV captures the context of different signals in the input message data, and hence has a vector form. Each value in the MCV can be thought of as a point in an n-dimensional space that contains the context of the series of signal values given as input. The MCV is fed into a GRU based decoder, which is then followed by a linear layer to reconstruct the input time-series of CAN message data with individual signal level values. Mean square error (MSE) is used to compute the loss between the input and the reconstructed input. Weights are updated using backpropagation through time. We design a recurrent autoencoder model for each message ID.

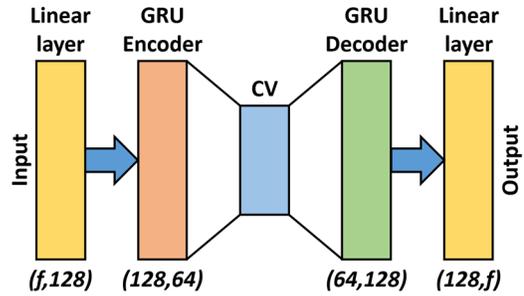

Figure 9: Proposed recurrent autoencoder network ($f$ is number of features i.e., number of signals in the input CAN message, MCV is message context vector)

#### 2) Training Process

The training process begins with the pre-processing of the CAN message data. Each sample in the dataset consists of a message ID and corresponding values of the signals within that message ID. The signal values are scaled between 0 to 1 for each signal type, as the range of signal values can be very large in some cases. Using such unscaled inputs can result in an extremely slow or very unstable training process. Moreover, as our goal is to reconstruct the input, scaling signal values also helps us avoid the problem of exploding gradients.

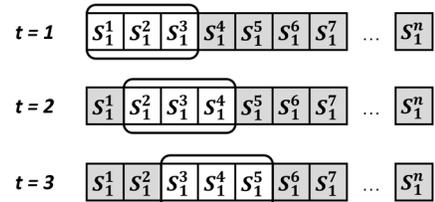

Figure 10: Rolling window based approach

After pre-processing the available data for training, it is split into training data (85%) and validation data (15%), and is prepared for training using a rolling window based approach. This involves selecting a window of fixed size and rolling it to the right by one time sample every time step. A rolling window size of three samples for three time steps is illustrated in Fig. 10, where the term $S_i^j$ represents the $i^{th}$ signal value at $j^{th}$ sample. The elements in the rolling window are collectively called as a subsequence and the subsequence size is equal to the size of the rolling window. As each subsequence consists of a set of signal values over time, the proposed recurrent autoencoder model tries to learn the temporal relationships that exist between the series of signal values. These signal level temporal relationships help in identifying more complex attacks such as continuous and playback (as discussed in section IV-C). The process of training using subsequences is done iteratively until the end of the sequence in training data.

During training, each iteration consists of a forward pass and a backward pass using backpropagation through time to update the weights and biases of the neurons (discussed in section III) based on the error value. At the end of the training, the model's learning is evaluated (forward pass only) using the validation data, which was not seen during the training. By the end of validation, the model has seen the full dataset once and this is known as an epoch. The model is trained for multiple epochs until the model reaches convergence. Moreover, the process of training and validation using subsequences is sped up by training the input data in groups subsequences known as mini-batches. Each mini-batch consists of multiple consecutive subsequences that are given to the model in parallel. The size of each mini-batch is commonly called batch size and it is a common practice to choose the batch size as a power of two. Lastly, to control the rate of update of parameters during backpropagation, a learning rate needs to be specified to the model. These hyperparameters such as subsequence size, batch size, learning rate, etc., are presented later in section VI-A.

### B. Inference and Detection

At runtime, the trained model is set to evaluation mode, meaning only the forward passes occur and the weights are not updated. In this phase, we can test for multiple attack scenarios (mentioned in Section IV-C), by simulating appropriate attack condition in the CAN message dataset.

Every data sample that goes through the model gets reconstructed and the reconstruction loss is sent to the detection module to compute a metric called *intrusion score* (IS). The IS helps us in identifying whether a signal is malicious or normal. We compute IS at a signal level to predict the signal that is under attack. The IS is computed at every iteration during inference, as a *squared error* to estimate the prediction deviation from the input signal value, as shown in (2).

$$IS_i = \left(\left(S_i^j - \hat{S}_i^j\right)^2\right) \quad \forall\ i \in [1, m] \quad (2)$$

where, $S_i^j$ represents $i^{th}$ signal value at $j^{th}$ sample, $\hat{S}_i^j$ denotes its reconstruction, and $m$ is the number of signals in the message. We observe a large deviation for predicted value from the input signal value (i.e., large IS value), when the signal pattern is not seen during the training phase, and a minimal IS value otherwise. This is the basis for our detection phase.

As we do not have a signal level intrusion label information in the dataset, we combine the signal level IS information into a message-level IS, by taking the maximum IS of the signals in that message as shown in (3).

$$MIS = \max(IS_1, IS_2 \ldots, IS_m) \quad (3)$$

In order to get adequate detection accuracy, the intrusion threshold (IT) for flagging messages needs to be selected carefully. We explored multiple choices for IT, using the best model from the training process. The best model is defined as the model with the lowest validation running loss during the training process. From this model, we log multiple metrics such as maximum, mean, median, 99.99%, 99.9%, 99% and 90% validation loss across all iterations as the choices for the IT. The analysis of IT metrics is presented in section VI-B.

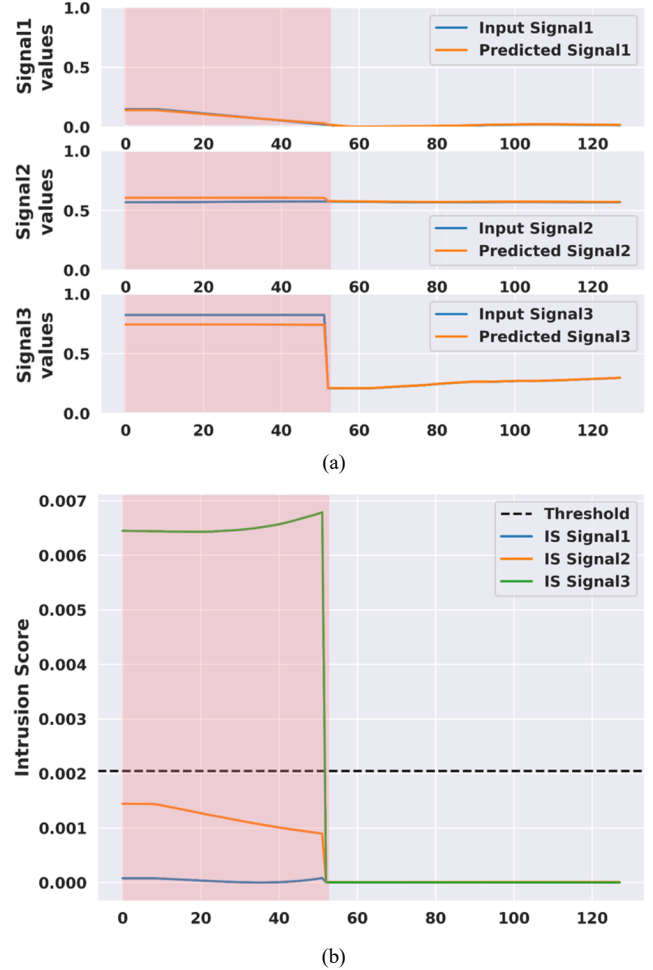

Figure 11: Snapshot of our proposed IDS checking a message with three signals under a plateau attack, where (a) shows the signal comparisons and (b) shows IS for signals and IS for the message and Intrusion flag

A snapshot of our IDS working in an environment with attacks is illustrated in Fig. 11a and 11b, with a plateau attack on a message with three signals, between time 0 and 50. Fig.11a shows the input (true) vs IDS predicted signal value comparisons for 3 signals. The red highlighted area represents the attack interval. It can be seen that for most of the time, the reconstruction is close for almost all signals except during the attack interval. Signal 3 is subjected to a plateau attack where the attacker held a constant value until the end of attack interval as shown in the third subplot of Fig. 11a (note the larger difference between the predicted and actual input signal values in that subplot, compared to for signals 1 and 2). Fig.11b shows the different signal intrusion scores for the 3 signals. The dotted black line is the intrusion threshold (IT). As mentioned earlier, the maximum of signal intrusion scores is chosen as message intrusion score (MIS), which in this case is the IS of signal 3. It can be observed from Fig. 11b that the intrusion score of signal 3 is above the IT, for the entire duration of the attack interval, highlighting the ability of *INDRA* to detect such attacks. The value of IT (equal to 0.002) in Fig. 11b is computed using the method discussed in section VI-B. Note that this value is specific to the example case shown in Fig. 11, and is not the threshold value used for our remaining experiments. Section VI.C describes how we select the IT value for our framework.

## VI. EXPERIMENTS

### A. Experimental Setup

To evaluate the performance of the *INDRA* framework, we first present an analysis for the selection of intrusion threshold (IT). Using the derived IT, we contrast it against the two variants of the same framework: *INDRA*-LED and *INDRA*-LD. The former removes the linear layer before the output and essentially leaving the GRU to decode the context vector. The term LED implies, (L)linear layer, (E) encoder GRU and (D) decoder GRU. The second variation replaces the GRU and the linear layer at the decoder with a series of linear layers (LD implies linear decoder). These experiments were conducted to test the importance of different layers in the network. However, the encoder end of the network is not changed because we require a sequence model to generate an encoding of the time-series data. We explored other variants as well, but they are not included in the discussion as their performance was poor compared to the LED and LD variants.

Subsequently, we compare the best variant of our framework with three prior works: Predictor LSTM (PLSTM [25]), Replicator Neural Network (RepNet [26]), and CANet [23]. The first comparison work (PLSTM) uses an LSTM based network that is trained to predict the signal values in the next message transmission. PLSTM achieves this by taking the 64-bit CAN message payload as the input, and learns to predict the signal at a bit-level granularity by minimizing the prediction loss. A log loss or binary cross-entropy loss function is used to monitor the bit level deviations between the real next signal values and the predicted next signal values, and the gradient of this loss function is computed using backpropagation to update the weights in the network. During runtime, PLSTM uses the prediction loss value to decide if that particular message is malicious or not. The second comparison work (RepNet) uses a series of RNN layers to increase the dimensionality of the input data and reconstruct the signal values by reducing back to the original dimensionality. RepNet achieves this by minimizing the mean squared error between the input and the reconstructed signal values. At runtime, large deviations between the input received signal and the reconstructed signal values are used to detect intrusions. Lastly, CANet unifies multiple LSTMs and linear layers in an autoencoder architecture and uses a quadratic loss function to minimize the signal reconstruction error. All experiments conducted with the *INDRA* variations and prior works are discussed in further subsections.

To evaluate our proposed framework with its variants and against prior works we used the SynCAN dataset that was developed by ETAS and Robert Bosch GmbH [23]. The dataset consists of CAN message data for 10 different IDs modeled after real-world CAN message data. The dataset comes with both training and test data with multiple attacks as discussed in section IV-C. Each row in the dataset consists of a timestamp, message ID, and individual signal values. Additionally, there is a label column in the test data with either 0 or 1 values indicating normal or malicious messages. The label information is available on a per message basis and does not indicate which signal within the message is subjected to the attack. We use this label information to evaluate our proposed IDS over several metrics such as detection accuracy and false positive rate, as is discussed in detail in the next subsections. Moreover, to simulate a more realistic attack scenario in the in-vehicle networks, the test data has normal CAN traffic between the attack injections. <u>Note</u>: We do not use the label information in the training data when training our model, as our model learns the patterns in the input data in an unsupervised manner.

All the machine learning based frameworks (*INDRA* and its variants, and comparison works) are implemented using Pytorch 1.4. We conducted several experiments to select the best performing model hyperparameters (number of layers, hidden unit sizes, and activation functions). The final model discussed in section V-A was trained using the SynCAN data set by splitting 85% of train data for training and the remaining for validation. The validation data is mainly used to evaluate the performance of the model at the end of every epoch. We trained the model for 500 epochs, using a rolling window approach (as discussed in section V-A.2) with the subsequence size of 20 messages and the batch size of 128. We also implemented an early stopping mechanism that monitors the validation loss across epochs and stops the training process if there is no improvement after 10 (patience) epochs. We chose the initial learning rate as 0.0001, and apply tanh activations after each linear and GRU layers. Moreover, we used the ADAM optimizer with the mean squared error (MSE) as the loss criterion. During testing, we used the trained model parameters and considered multiple test data inputs to simulate attack scenarios. We monitored the intrusion score metric (as described in section V-C) and the computed intrusion threshold to flag the message as malicious or normal. We computed several performance metrics such as detection accuracy, false positives, etc. to evaluate the performance of our model. All the simulations are run on an AMD Ryzen 9 3900X server with an Nvidia GeForce RTX 2080Ti GPU.

Lastly, before showing the experimental results, we present the following definitions in the context of IDS:
- *True Positive (TP)*- when the IDS detects an actual malicious message as malicious;
- *False Negative (FN)*- when the IDS detects an actual malicious message as normal;
- *False Positive (FP)*- when the IDS detects a normal message as malicious (aka false alarm);
- *True Negative (TN)*- when the IDS detects an actual normal message as normal.

We focus on two key performance metrics: *(i) Detection accuracy*- a measure of IDS ability to detect intrusions correctly, and *(ii) False positive rate*: also known as false alarm rate. These metrics are calculated as shown in (4) and (5):

$$Detection\ Accuracy = \frac{TP+TN}{TP+FN+FP+TN} \quad (4)$$

$$False\ Positive\ Rate = \frac{FP}{FP+TN} \quad (5)$$

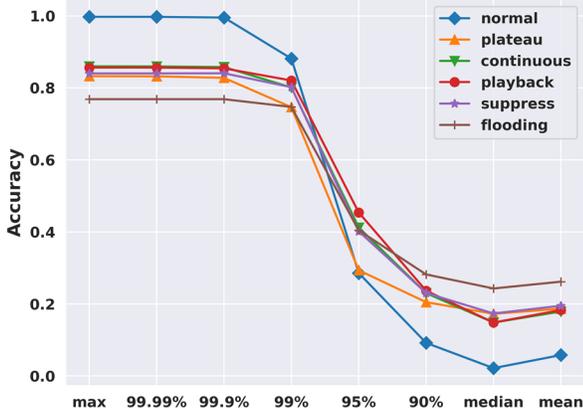

(a)

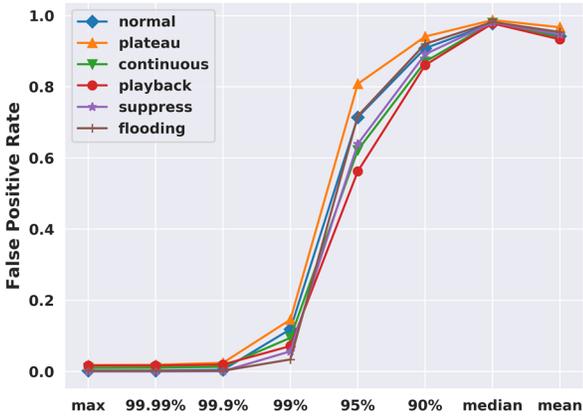

(b)

Figure 12: Comparison of (a) detection accuracy, and (b) false positive rate for various candidate options of intrusion threshold (IT) as a function of validation loss under different attack scenarios. (% refers to percentile not percentage)

### B. Intrusion Threshold Selection

In this subsection, we present an analysis for the selection of intrusion threshold (IT) by considering various options such as max, median, mean, and different quantile bins of validation loss of the final model. The model reconstruction error for the normal message should be much smaller than the error for malicious messages. Hence, we want to explore several candidate options to achieve this goal, that would work across multiple attack and no-attack scenarios. Having a large threshold value can make it harder for the model to detect the attacks that change the input pattern minimally (e.g., continuous attack). On the other hand, having a small threshold value can trigger multiple false alarms, which is highly undesirable. Hence it is important to select an appropriate threshold value to optimize the performance of the model.

Fig. 12a and 12b illustrate the detection accuracy and false positive rate respectively for various candidate options to calculate IT, under different attack scenario. It is clear from the results in the figure that selecting higher validation loss as the IT can lead to a high accuracy and low false alarm rate. However, choosing a very high value (e.g., 'max' or '99.99 percentile') can sometimes result in missing small variations in the input patterns that are found in more sophisticated attacks. From our experiments we found the maximum and 99.99 percentile values to be very close. In order to capture the attacks that produce small deviations, we selected a slightly smaller threshold that would still perform similar to max and 99.99 percentile thresholds on all of our current attack scenarios. Hence, in this work, we choose the 99.9th percentile value of the validation loss as the value of the intrusion threshold (IT). We use the same IT value for the remainder of the experiments discussed in the next subsections.

### C. Comparison of INDRA Variants

After selecting the correct intrusion threshold from the previous subsection, we use that criterion and evaluate our proposed *INDRA* framework with two other variants: *INDRA*-LED, and *INDRA*-LD. The main intuition behind evaluating different variants of *INDRA* is to analyze the impact of different types of layers in the model on the performance metrics discussed in section VI-A.

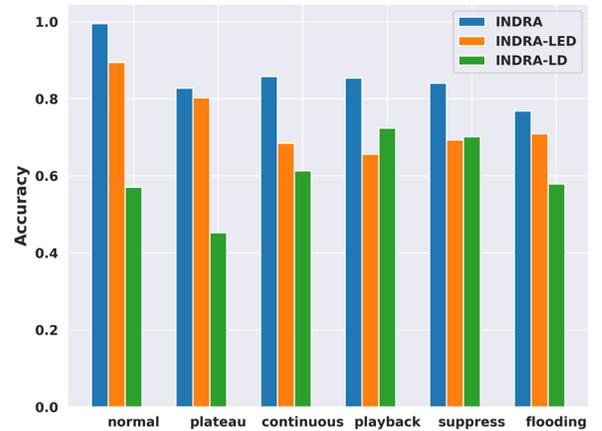

(a)

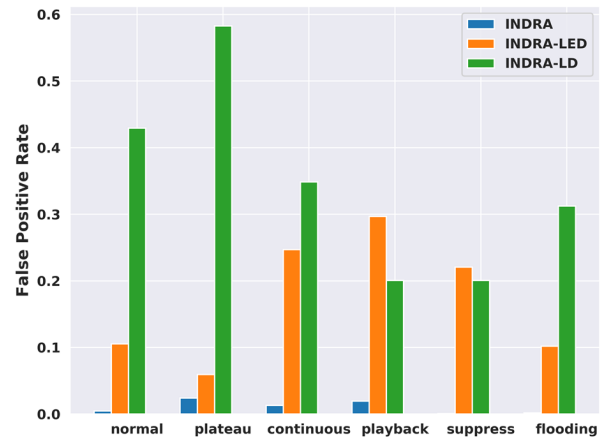

(b)

Figure 13: Comparison of (a) detection accuracy, and (b) false positive rate for INDRA and its variants INDRA-LED and INDRA-LD under different attack scenarios.

Fig. 13a shows the detection accuracy for our *INDRA* framework and its variants on y-axis with different attack types and for a no-attack scenario (normal) on the x-axis. We can observe that *INDRA* outperforms the other two variants and has high accuracy in most of the attack scenarios. It is to be noted that the high accuracy is achieved by monitoring at a signal level unlike prior works that monitor at the message level.

The false positive rate or false alarm rate of *INDRA* and other variants under different attack scenarios is shown in Fig. 13b. It is evident that *INDRA* has the lowest false positive rate and highest detection accuracy compared to the other variants. Moreover, *INDRA*-LED which is just short of a linear layer at the decoder end is the second best performing model after *INDRA*. *INDRA*-LED's ability to use a GRU based decoder helps in reconstructing the MCV back to original signals. It can be clearly seen in both Fig. 13a and 13b, that the lack of GRU layers on the output decoder end for *INDRA*-LD leads to a significant performance degradation. Hence, we chose *INDRA* as our candidate model for subsequent experiments.

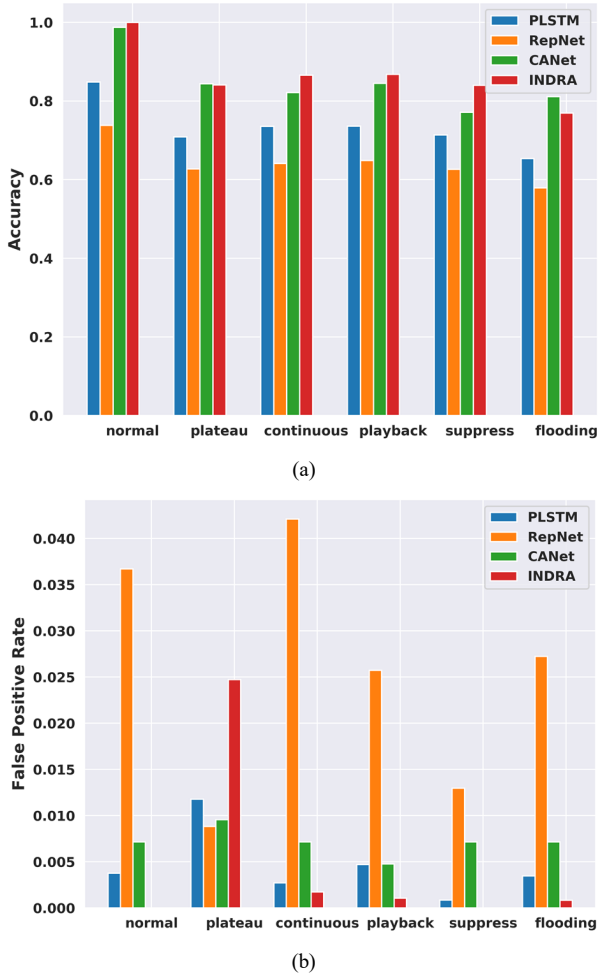

(a)

(b)

Figure 14. Comparison of (a) detection accuracy, and (b) false positive rate of *INDRA* and the prior works PLSTM [25], RepNet [26] and CANet [23].

### D. Comparison with Prior Works

We compare our *INDRA* framework with PLSTM [25], RepNet [26] and CANet [23], which are some of the best known prior works in the IDS area. Fig. 14a and 14b show the detection accuracy and false positive rate respectively for the various techniques under different attack scenarios.

From the results shown in Fig. 14, it is evident that *INDRA* achieves high accuracy for each attack scenario and also achieves low positive rates for most of the scenarios. The ability to monitor signal level variations along with the more cautious selection of intrusion threshold gives *INDRA* an advantage over comparison works. PLSTM and RepNet use the maximum validation loss in the final model as the threshold to detect intrusions in the system, while CANet uses interval based monitoring to detect attacks. Selecting a larger threshold helped PLSTM to achieve slightly lower false positive rates for some scenarios, but it hurt the ability of both PLSTM and RepNet to detect attacks with small variations in the input data. This is because the deviations produced by some of the complex attacks are small and due to the large thresholds the attacks go undetected. Moreover, the interval based monitoring in CANet struggles with finding an optimal value for the thresholds. Lastly, the false positive rates of *INDRA* are still significantly low with the maximum of 2.5% for plateau attacks. Please note that the y-axis in Fig. 14b has a much smaller scale than in Fig. 14a, and the magnitude of the false positive rate is very small.

### E. IDS Overhead Analysis

In this subsection, we present a detailed analysis of the overhead incurred by our proposed IDS. We quantify the overhead in terms of both memory footprint and time taken to process an incoming message i.e., inference time. The former metric is important as the resource constrained automotive ECUs have limited available memory, and it is crucial to have a low memory overhead to avoid interference with real-time automotive applications. The inference time not only provides important information about the time taken to detect the attacks, but also can be used to compute the utilization overhead on the ECU. Thus, we choose the abovementioned two metrics to analyze the overhead and quantify the lightweight nature of our proposed IDS.

To accurately capture the overhead of our proposed *INDRA* framework and the prior works, we implemented our proposed IDS approach on an ARM Cortex- A57 CPU on a Jetson TX2 board, which has similar specifications to the state-of-the-art multi-core ECUs. Table II presents the memory footprint of our proposed *INDRA* framework and the prior works mentioned in the previous subsections. It is clear that our proposed *INDRA* framework has a low memory footprint compared to the prior works, except for the RepNet [26]. However, it is important to observe that even though our proposed framework has slightly higher memory footprint compared to the RepNet [26], we outperform all of the prior works including RepNet [26] in all performance metrics under different attack scenarios, as shown in Fig. 14. The heavier (high memory footprint) models can provide the ability to capture a large variety of details about the system behavior, but they are not an ideal choice for resource constrained automotive systems. On the other hand, a much lighter model such as RepNet, fails to capture key details about the system behavior due to its limited parameters and therefore suffers from performance issues.

TABLE II
MEMORY FOOTPRINT COMPARISON BETWEEN INDRA FRAMEWORK AND
THE PRIOR WORKS PLSTM [25], REPNET [26] AND CANET [23]

| Framework | Memory footprint (KB) |
|---|---|
| PLSTM [25] | 13,417 |
| RepNet [26] | 55 |
| CANet [23] | 8,718 |
| INDRA | 443 |

TABLE III
INFERENCE TIME COMPARISONS BETWEEN INDRA FRAMEWORK AND
THE PRIOR WORKS PLSTM [25], REPNET [26] AND CANET [23] USING
SINGLE CORE, AND DUAL CORE CONFIGURATIONS

| Framework | Average inference time (μs) | |
|---|---|---|
| | Single core ARM Cortex A57 CPU | Dual core ARM Cortex A57 CPU |
| PLSTM [25] | 681.18 | 644.76 |
| RepNet [26] | 19.46 | 21.46 |
| CANet [23] | 395.63 | 378.72 |
| INDRA | 80.35 | 72.91 |

In order to understand the inference overhead, we benchmarked the different IDS frameworks on an ARM Cortex-A57 CPU. In this experiment, we consider different system configurations to encompass a wide variety of ECU hardware that is available in the state-of-the-art vehicles. Based on the available hardware resources on the Jetson TX2, we selected two different system configurations. The first configuration utilizes only one CPU core (single core), while the second configuration uses two CPU cores.

We ran the frameworks 10 times for the different CPU configurations and computed the average inference time (in μs), as shown in Table III. From the results in table III, it is clear that our proposed *INDRA* framework has significantly faster inference times compared to the prior works (excluding RepNet) under all three configurations. This is partly due to the lower memory footprint of our proposed IDS. As mentioned earlier, even though RepNet has a lower inference time, it has the worst performance out of all frameworks, as shown in Fig. 14. The large inference times for the better performing frameworks can impact the real-time performance of the control systems in the vehicle, and can result in catastrophic missing of deadlines. We also believe that using a dedicated deep learning accelerator (DLA) would give us significant speed up compared to the above presented configurations.

Thus, from Fig. 14, and table II and table III, it is clear that INDRA achieves a clear balance of having superior intrusion detection performance while maintaining low memory footprint and fast inference times, making it a powerful and lightweight IDS solution.

*F. Scalability Results*

In this subsection we present an analysis on the scalability of our proposed IDS by studying the system performance using the ECU utilization metric as a function of increasing system complexity (number of ECUs and messages).

Each ECU in our system model has a real-time utilization ($U_{RT}$) and an IDS utilization ($U_{IDS}$) from running real-time and IDS applications respectively. In this work, we primarily focus on analyzing the IDS overhead ($U_{IDS}$), as it is a measure of the compute efficiency of the IDS. Since the safety-critical messages monitored by the IDS are periodic in nature, the IDS can be modeled as a periodic application with period that is the same as the message period [32]. Thus, monitoring an $i^{th}$ message $m_i$ results in an induced IDS utilization ($U_{IDS, mi}$) at an ECU, and can be computed as:

$$U_{IDS,m_i} = \left(\frac{T_{IDS}}{P_{m_i}}\right) \quad (6)$$

where, $T_{IDS}$ and $P_{mi}$ indicate the time taken by the IDS to process one message (inference time), and the period of the monitored message, respectively. Moreover, the sum of all IDS utilizations as a result of monitoring different messages is the overall IDS utilization at that ECU ($U_{IDS}$) and is given by:

$$U_{IDS} = \sum_{i=1}^{n} U_{IDS,m_i} \quad (7)$$

To evaluate the scalability of our proposed IDS, we consider six different system sizes. Moreover, we consider a pool of commonly used message periods {1, 5, 10, 15, 20, 25, 30, 45, 50, 100} (all periods in ms) in automotive systems to sample uniformly, when assigning periods to the messages in the system. These messages are evenly distributed among different ECUs and the IDS utilization is computed using (6) and (7). In this work, we assume a pessimistic scenario where all of the ECUs in the system have only a single core. This would allow us to analyze the worst case overhead of the IDS.

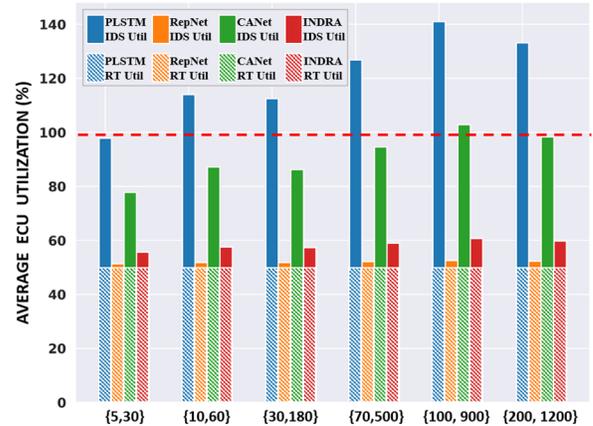

Fig. 15 Scalability results of our proposed IDS for different system sizes and the prior works PLSTM [25], RepNet [26] and CANet [23]

Figure 15 illustrates the average ECU utilization under various system sizes denoted by {$p$, $q$}, where $p$ is the number of ECUs and $q$ is the number of messages in the system. A very pessimistic estimate of 50% real-time ECU utilization for real-time automotive applications is assumed ("RT Util", as shown in the dotted bars). The solid bars on top of the dotted bars represent the overhead incurred by the IDS executing on the ECUs, and the red horizontal dotted line represents the 100% ECU utilization mark. It is important to avoid exceeding the 100% ECU utilization under any scenario, as it could induce undesired latencies that could result in missing deadlines for time-critical automotive applications, which can be catastrophic. It is clear from the results that the prior works

PLSTM and CANet incur heavy overhead on the ECUs while RepNet and our proposed *INDRA* framework have very minimal overhead that scales favorably to increasing system sizes. From the results in this section (Figures 14, 15; Tables II, III), it is apparent that not only does INDRA achieve better performance in terms of both accuracy and low false positive rate for intrusion detection than state-of-the-art prior work, but it is also lightweight and scalable.

## VII. Conclusion

In this paper, we proposed a novel recurrent autoencoder based lightweight intrusion detection system called INDRA for distributed automotive embedded systems. We proposed a metric called the intrusion score (IS), which measures the deviation of the prediction signal from the actual input. We also presented a thorough analysis of our intrusion threshold selection process and compared our approach with the best known prior works in this area. The promising results indicate a compelling potential for utilizing our proposed approach in emerging automotive platforms. In our future work, we plan to exploit the dependencies that exist between signals to improve the performance of our intrusion detection framework.

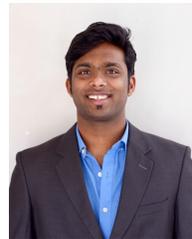

**Vipin Kumar Kukkala** (S'13) received his B. Tech degree in electronics and communications engineering from Jawaharlal Nehru Technological University, Hyderabad, India in 2013. He is currently pursuing the Ph.D. degree in computer engineering at Colorado State University, USA. His current research interests include design of next-gen automotive networks, and security in cyber-physical systems.

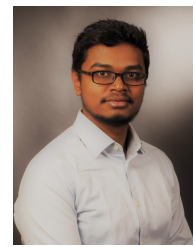

**Sooryaa Vignesh Thiruloga** (S'20) received B. Tech degree in electronics and communications engineering from Amrita Vishwa Vidhyapeetham in 2019. He is currently pursuing the M.S. degree in computer engineering at Colorado State University, USA with focus on automotive network security.

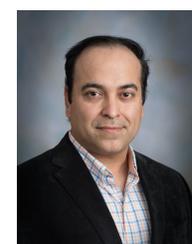

**Sudeep Pasricha** (M'02, SM'13) received his Ph.D. degree in computer science from the University of California at Irvine, USA, in 2008. He is currently a Professor and Chair of Computer Engineering in the Department of Electrical and Computer Engineering at Colorado State University. His current research interests include hardware-software co-design for cyber-physical systems, and optimizations for energy, reliability, and security in manycore embedded systems.